\documentclass[prd,twocolumn,amsmath,amssymb,showpacs,floatfix,superscriptaddress,nofootinbib,preprintnumbers]{revtex4}

\usepackage{epsfig}
\usepackage{bm}
\usepackage{latexsym}
\usepackage{natbib}
\usepackage{url}
\usepackage{dcolumn}
\usepackage{color}
\usepackage[usenames,dvipsnames]{xcolor}
\usepackage{amsfonts,amssymb,amsmath}
\usepackage{graphicx,epsfig}
\usepackage{psfrag}
\usepackage{subfigure}
\usepackage{comment}
\usepackage{lipsum}
\usepackage{caption}
\usepackage{blindtext}
\usepackage{multirow}
\usepackage{hyperref}
\hypersetup{colorlinks=true}

\def\be{\begin{equation}}
\def\ee{\end{equation}}
\def\gev{{\rm \,Ge\kern-0.125em V}}

\begin{document}
 
%---------------------------------------------------------------------%
\title{On signatures of spontaneous collapse dynamics modified single field inflation}
%---------------------------------------------------------------------%

\author{Shreya Banerjee}
 \email{shreya.banerjee@mailhost.tifr.res.in}
\affiliation{Tata Institute of Fundamental Research, Homi Bhabha Road, Mumbai 400008, India}

\author{Suratna Das}
 \email{suratna@iitk.ac.in}
\affiliation{Indian Institute of Technology, Kanpur 208016, India}

\author{K. Sravan Kumar}
 \email{sravan@ubi.pt}
\affiliation{Departmento de F\'{i}sica and Centro de Matem\'atica e Aplica\c{c}\~oes (CMA-UBI), Universidade da Beira Interior, Rua Marqu\^es D'\'Avila e Bolama,
6200 Covilh\~a, Portugal}

\author{T. P. Singh}
 \email{tpsingh@mailhost.tifr.res.in}
\affiliation{Tata Institute of Fundamental Research, Homi Bhabha Road, Mumbai 400008, India}

\date{\today}

\begin{abstract}
The observed classicality of primordial perturbations, despite their quantum origin during inflation, calls for a mechanism for quantum-to-classical transition of these initial fluctuations. As literature suggests a number of plausible mechanisms which try to address this issue, it is of importance to seek concrete observational signatures of the various approaches in order to have a better understanding of the early universe dynamics. Among these several approaches, it is the spontaneous collapse dynamics of Quantum Mechanics  which is most viable for leaving discrete observational signatures as collapse mechanism inherently changes the generic quantum dynamics. We observe in this study that the observables from the scalar sector, i.e. scalar tilt $n_s$, running of scalar tilt $\alpha_s$ and running of running of scalar tilt $\beta_s$, can not potentially distinguish a collapse modified inflationary dynamics in the realm of canonical scalar field and $k-$inflationary scenarios. The only distinguishable imprint of collapse mechanism lies in the observables of tensor sector in the form of modified consistency relation and a blue-tilted tensor spectrum only when the collapse parameter $\delta$ is non-zero and positive. 
\end{abstract}


\maketitle

%-----------------------------------------------------------------------%
%=======================================================================================
 \section{Introduction}

The advent of precision cosmology has rendered the paradigm of cosmic inflation \cite{Starobinsky:1980te, Guth:1980zm, Linde:1981mu} impressively successful observationally as many of its non-trivial predictions, like near scale-invariance of inflationary scalar perturbations, their Gaussian and adiabatic nature and smaller tensor perturbations than scalar, have stood the test of time. The mechanism of cosmic inflation describes the evolution of primordial quantum perturbations on a gravitational background which eventually give rise to the large scale structures (LSS) and the cosmic microwave background (CMB) anisotropies. These LSS and the CMB anisotropies are then being probed observationally to put to test the dynamics of cosmic inflation.  As these are classical observables (the LSS and the CMB anisotropies) which are being probed to verify their quantum origin, validity of such a paradigm calls for a quantum-to-classical transition mechanism for these primordial quantum perturbations before they seed the observable classical structures \cite{Polarski:1995jg}. 

This leads us to an equivalent problem in laboratory systems, known as the measurement problem of Quantum Mechanics. Copenhagen Interpretation of Quantum Mechanics, first proposed by Niels Bohr in 1920s, is the simplest one to deal with such a fundamental problem, which states that the process of measurement yields the collapse of the wavefunction of otherwise superposed quantum states. But such an interpretation falls short in a cosmological setup, like inflationary mechanism, where the very process of measurement is ill-defined. To go beyond the Copenhagen Interpretation of Quantum Mechanics, one can broadly classify the various existing approaches into two categories : 1. mechanisms which do not modify the inherent dynamics of Quantum Mechanics, such as decoherence \cite{Schlosshauer:2003zy} along with many-worlds interpretation \cite{1957RvMP...29..454E} and Bohmian mechanics \cite{1952PhRv...85..166B}, and, 2. mechanisms which modify the standard dynamics of Quantum Mechanics, such as the spontaneous collapse dynamics \cite{Bassi:2003gd, Bassi:2012bg}. 

All these approaches to address the measurement problem of Quantum Mechanics in laboratory systems have been studied in the context of inflationary dynamics. Hence inflationary paradigm can act as the testing ground of all these approaches if each of them leaves distinct signatures on the primordial observables. It is known that though quantum decoherence successfully suppresses the off-diagonal components of the reduced density matrix, generating classical stochastic perturbations from inflation, it does not leave any observable signature which can be experimentally probed \cite{Kiefer:2008ku, Lombardo:2005iz, Martineau:2006ki, Prokopec:2006fc, Nelson:2016kjm}. Besides, decoherence, by construction, does not resolve the issue of single outcome and appeals for the many-worlds interpretation which is again observationally non-falsifiable. On the other hand, though Bohmian mechanics are not expected to leave any imprint on observables, but because of non-equilibrium initial conditions on the hidden sector variables, such mechanisms leave imprints on the CMB, explaining the low-power anomaly of CMB at large angular scales \cite{Valentini:2008dq}. The aim of this paper is to quantify the observable signature of collapse dynamics while applying it to cosmic inflation.

As mentioned above, spontaneous collapse dynamics tends to modify the generic quantum dynamics and hence is most viable for leaving observational imprints both in laboratory systems and in cosmological setup. In literature, there are mainly two approaches to incorporate collapse dynamics in inflationary mechanism. In one approach, developed by Sudarsky and collaborators, collapse dynamics is incorporated to generate the primordial perturbations in an otherwise homogeneous and isotropic background \cite{Leon:2011ca, PhysRevD.87.104024}. Such incorporation of collapse dynamics in inflationary mechanism yields the standard Harrison-Zel'dovich spectrum while yielding no or highly suppressed tensor modes, which are both in good agreement with present observations. The second approach \cite{Martin:2012pea, Das:2013qwa} treats the primordial quantum perturbations generic to the inflationary dynamics and seeks for their quantum-to-classical transition through incorporation of spontaneous collapse dynamics. In this work we will only consider this second approach. The observational consequences of applying such collapse mechanism (in the realm of a specific collapse model, dubbed as Continuous Spontaneous Localization (CSL) model \cite{PhysRevA.42.78}) to canonical single field slow-roll inflationary dynamics, where the inflaton field has canonical kinetic term and minimal coupling with gravity, has been studied in \cite{Das:2014ada} where it was shown that the collapse dynamics can significantly modify the scalar and tensor spectral indices $(n_s,\,n_T)$ and the consistency relation of the canonical single-scalar field inflationary model ($r=-8n_T$, $r$ being the tensor-to-scalar ratio) while leaving the running and the running of the running of scalar and tensor spectral indices unaltered. Hence it was argued that accurate measurements of both scalar and tensor tilts along with their running and running of running can potentially distinguish this collapse-dynamics modified canonical single field slow-roll scenario from the canonical case along with other scenarios which tend to modify the single-field consistency relation (such as curvaton scenario \cite{Fujita:2014iaa}, multifield scenario \cite{Kim:2006ys}, non-Bunch Davies initial vacuum \cite{Ashoorioon:2014nta}). We note that though the collapse-modified inflationary dynamics has three free parameters $\tilde\gamma_0$, $\alpha$ and $\beta$, the observables of the canonical inflationary model get modified by one parameter $\delta\,\,(\equiv 3+\alpha-\beta)$. As one is yet to develop a formal field theoretic version of CSL model, the values of the collapse parameters $\alpha$ and $\beta$ can be chosen phenomenologically. While $1<\alpha<2$ is required to explain the classicalization of the modes, there is no bound on $\beta$ and it is expected that the parameter $\delta$ can only be of the order of slow-roll parameters in order to render the scenario observationally viable. Hence it is always possible that $\delta$ is identically zero (when $\beta=3+\alpha$) in which case the collapse dynamics would leave no imprint on the observables and would render this scenario observationally identical to the generic one.

%The simplest realization of inflationary mechanism manifest itself in the form of single-field slow-roll scenario where the inflaton field has a canonical kinetic term and a minimal coupling with gravity. The inflationary era is driven by the flat potential of the inflaton field and the primordial perturbations are originated from the quantum fluctuation of the same inflaton field. This simplest scenario is in very good agreement with current data, though the confirmation of such a simple scenario still awaits the accurate measurement of the inflationary observables. This simplest scenario has been considered in  \cite{Martin:2012pea, Das:2013qwa, Das:2014ada} to analyze the quantum-to-classical transition of primordial perturbations according to spontaneous collapse dynamics. 

It is observed by  Mukhanov and collaborators \cite{ArmendarizPicon:1999rj, Garriga:1999vw} that it is possible to generalise this simplest canonical model of inflation by allowing non-canonical kinetic term of the inflaton field. Inflation, in such scenarios, can be driven by the non-canonical kinetic term of inflaton, and hence named as $k-$inflation, which avoids the requirement of the very flat inflaton potentials for slow-roll which is often difficult to achieve in realistic models. Also, non-canonical kinetic terms of scalar field often appear in String motivated models \cite{Alishahiha:2004eh, Gibbons:2002md} which makes this kind of realization more appealing. The main difference between the canonical scalar field and non-canonical $k-$inflation scenarios is the speed of sound $c_s$ at which the inflationary scalar perturbations propagate during inflation: in canonical inflation the scalar perturbations travel with $c_s=1$ (the speed of light) while in $k$-inflation they propagate with sound speed $c_s<1$ \footnote{$0<c_s<1$ is required to avoid instabilities and ultraluminous propagation speed of the primordial modes}. This difference manifests itself by modifying the consistency relation of the canonical single-field slow-roll model ($r=-8n_Tc_s$) which for $c_s=1$ reduces to the generic canonical case.\footnote{$k-$inflation also yields larger Bispectrum non-Gaussianity than the canonical case and thus puts an upper bound on $c_s$ \cite{Ade:2015lrj}. But as calculating non-Gaussianity with collapse dynamics becomes tricky due to non-linear dynamics of primordial modes, we will not try to address the issue of non-Gaussinity here and would defer the topic for a future study.} The tensor modes in both these scenarios travel with speed of light. 
We will consider here such generalised single-field inflationary  scenario to apply the CSL collapse dynamics and would investigate whether the collapse dynamics can leave any non-trivial imprint on observables, like scalar tilt $n_s$, tensor tilt $n_T$, tensor-to-scalar ratio $r$, running of scalar tilt $\alpha_s$ and running if running of scalar tilt $\beta_s$, (even in a case when $\delta=0$) which can potentially distinguish the collapse-modified inflationary dynamics from the generic case.

This article is organized as follows : In Section \ref{k-inf} we would briefly discuss the background and perturbation dynamics in generic $k-$inflation scenario. Section \ref{csl-inf} would contain brief discussion of how to implement collapse dynamics in a caonical single-field slow roll inflation. We will then implement the collapse mechanism in $k-$inflationary scenario in Section \ref{csl-kinf} and quantify the effects of collapse on inflationary observables. We will then discuss the observational consequences of collapse mechanism for single-field slow-roll inflation in Section \ref{discussion} and then we conclude in Section \ref{conclusion}.
%------------------------------------------------------------------------------%
\section{$k-$inflation in brief}
\label{k-inf}

In $k-$inflation \cite{ArmendarizPicon:1999rj, Garriga:1999vw} the action for the inflaton field considered is the most general action which involves the first derivatives of the field and can be written as 
\begin{eqnarray}
S=-\frac{1}{16\pi G}\int d^4x \sqrt{-g}R+\int d^4x\sqrt{-g}p(X,\varphi),
\end{eqnarray}
where $X\equiv\frac12g^{\mu\nu}\partial_\mu\varphi\partial_\nu\varphi$ is the canonical kinetic term of inflaton. We note that varying the matter Lagrangian $p(X,\varphi)$ with respect to the metric one gets the energy momentum tensor of the inflaton field as 
\begin{eqnarray}
T_{\mu\nu}=(\varepsilon+p)u_\mu u_\nu-pg_{\mu\nu},
\end{eqnarray}
where the energy density is $\varepsilon=2Xp_{,X}-p$, and the pressure is $p=p(X,\varphi)$ which is the Lagrangian of the scalar field. Here $u_\mu\equiv \varphi_\mu/(1+X)^{1/2}$.

Considering a flat Friedmann-Robertson-Walker metric as the background geometry, the above Lagrangian yields the standard Friedmann equations and continuity equation as
\begin{eqnarray}
H^2&=&\frac{1}{3M_{\rm Pl}^2}\varepsilon,\nonumber\\
\dot H&=&-\frac{1}{M_{\rm Pl}^2}(\varepsilon+p),\nonumber\\
\dot\varepsilon&=&-3H(\varepsilon+p),
\end{eqnarray}
where $H$ is the Hubble parameter and overdot represents derivative with respect to cosmic time $t$ and $M_{\rm Pl}\equiv (8\pi G)^{-1}$ is the reduced Planck mass. These equations yield 
\begin{eqnarray}
\dot p=-3c_s^2H(\varepsilon+p)+\dot\varphi(p_{,\varphi}-c_s^2\varepsilon_{,\varphi}),
\end{eqnarray}
where the parameter 
\begin{eqnarray}
c_s^2\equiv\frac{p_{,X}}{\varepsilon_{,X}}=\frac{\varepsilon+p}{2X\varepsilon_{,X}}
\end{eqnarray}
turns out to be the `speed of sound' for inflationary perturbations. 

One of the main motivations of proposing $k-$inflation was to have a possibility of implementing inflation even when the potential of the inflaton field either tends to zero or grows very fast, due to radiative corrections, barring slow-roll. Hence, in absence of any potential term, one would expect the Lagrangian of $k-$inflation $p(\varphi,X)$ to vanish when $X\rightarrow0$. In the region $X=0$ one thus can expand the Lagrangian as 
\begin{eqnarray}
p(\varphi,X)=K(\varphi)X+L(\varphi)X^2+\cdots.
\end{eqnarray}
It is noted in \cite{ArmendarizPicon:1999rj} that when the Lagrangian is a function of pure kinetic term $X$ then the evolution equation tends towards an attractor $p=-\varepsilon$, which then leads to a pure exponential expansion of universe required to drive inflation. But such simplification leads to a obvious problem of graceful exit. Hence to implement slow-roll, in order to avoid the graceful exit problem, one must have the general case of the Lagrangian $p(\varphi,X)$ and in the $X\rightarrow 0$ limit one gets :
\begin{eqnarray}
p(\varphi,X)&=&K(\varphi)X+L(\varphi)X^2,\nonumber\\
\varepsilon(\varphi,X)&=&K(\varphi)X+3L(\varphi)X^2.
\end{eqnarray}
If both the coefficients $K(\varphi)$ and $L(\varphi)$ are positive, then this Lagrangian would not lead to an inflationary solution. 
In a case, where $K(\varphi)<0$ (dubbed as strong-coupling region), the above solutions tend towards an inflationary fixed point $p_{\rm fixed}=-\varepsilon_{\rm fixed}$ which can be seen as follows. At inflationary fixed point we have $\varepsilon+p=0$, implying $Xp_{,X}=0$. Without any loss of generality we can take $L(\varphi)=1$, which yields the zeroth order slow-roll quantities as $X_{\rm sr}=\frac12\tilde K(\varphi_{\rm sr})$, $\varepsilon_{\rm sr}=\frac14\tilde K^2(\varphi_{\rm sr})$ and $H_{\rm sr}=(2\sqrt{3}M_{\rm Pl})^{-1}\tilde K(\varphi_{\rm sr})$. The `slow-roll' conditions are maintained as long as the departure from these zeroth order slow-roll quantities are small, i.e. $\delta X/X_{\rm sr}\ll1$. In this case, where $L(\varphi)=1$, it yields $\partial(\tilde K)^{-1/2}/\partial\varphi\ll3/2$. The Hubble slow-roll parameters can be defined in the usual way for $k-$inflation scenario as 
\begin{eqnarray}
\epsilon_0=-\frac{\dot H}{H^2},\quad \epsilon_{n+1}=\frac{\dot\epsilon_n}{H\epsilon_n}.
\end{eqnarray}
Along with the standard Hubble slow-roll parameters we also define a series of slow-roll-like parameters related to the parameter $c_s$ as \cite{vandeBruck:2016rfv}
\begin{eqnarray}
s_0=\frac{\dot c_s}{Hc_s},\quad s_{n+1}=\frac{\dot s_n}{Hs_n},
\end{eqnarray}
and demand $\epsilon_n,\, s_n\ll1$ during slow-roll $k-$inflation. In $k-$inflationary scenario $\epsilon_0=(3/2)(1+p/\varepsilon)$, $\epsilon_1=H^{-1}\left(\ln(1+p/\varepsilon)\right)\dot{}$ and $s_0=H^{-1}(\ln(c_s))\dot{}$\,.

Inflationary perturbations in such a $k-$inflationary scenario can be fully described in terms of two scalar perturbations : the inflaton fluctuations $\delta\varphi$ and the metric perturbation $\Phi$, if one considers the longitudinal gauge of the perturbed metric. The advantage of this particular choice of gauge is that these scalar perturbations coincide with the gauge invariant quantities : gauge invariant inflaton perturbations and Bardeen potential respectively. One can construct another gauge-invariant scalar $\zeta$ out of $\delta\varphi$ and $\Phi$ as 
\begin{eqnarray}
\zeta=\Phi+H\frac{\delta\varphi}{\dot\varphi},
\end{eqnarray}
which turns out to be the same as the comoving curvature perturbation $\mathcal{R}$ in this gauge.
The first order action $\zeta$ which yields the correct equation of motion of the fields is 
\begin{eqnarray}
S=\frac{1}{2}\int z^2\left[\zeta'^2+c_s^2\zeta\partial_i\partial^i\zeta\right]d\tau d^3x,
\end{eqnarray}
where $\tau$ is the conformal time, prime denotes derivative with respect to $\tau$ and $z$ is defined as 
\begin{eqnarray}
z=\frac{a\sqrt{\varepsilon+p}}{c_sH}=\frac{aM_{\rm Pl}}{c_s}\sqrt{2\epsilon_0}\,.
\label{k-inf-z}
\end{eqnarray}
We can define a canonical quantization variable $v=z\zeta$ to rewrite the action as 
\begin{eqnarray}
S=\frac12\int\left[v{'^2}+c_s^2v\partial_i\partial^i v+\frac{z''}{z}v^2\right]d\tau d^3x,
\label{action-v}
\end{eqnarray} 
which yields the equation of motion of the variable $v$ in momentum space as
\begin{eqnarray}
v_k''+\left(c_s^2k^2-\frac{z''}{z}\right)v_k=0,
\label{v-eom}
\end{eqnarray}
where $v_k$ is the mode function defined as 
\begin{eqnarray}
v=\int\frac{d{\mathbf k}^3}{(2\pi)^3}\left[v_k(\tau)\hat a_{\mathbf k}e^{i{\mathbf k}\cdot{\mathbf x}}+v^*_k(\tau)\hat a^\dagger_{\mathbf k}e^{-i{\mathbf k}\cdot{\mathbf x}}\right].
\end{eqnarray}
Here $[\hat a_{\mathbf k},\hat a^\dagger_{\mathbf k'}]=(2\pi)^3\delta^3({\mathbf k}-{\mathbf k'})$.
We note that during slow-roll $z''/z\sim a''/a\sim 2(aH)^2$, and a mode $k$ crosses the `sound horizon' $(c_sH^{-1})$ when $k=aH/c_s$. Thus in the subhorizon limit $k\gg aH/c_s$ one can drop the potential term $z''/z$ and the solution of the above equation for positive frequency mode would be 
\begin{eqnarray}
v_k\sim \frac{e^{-ic_sk\tau}}{\sqrt{2c_sk}}.
\end{eqnarray}
When a mode is much beyond the `sound horizon' then the potential term $z''/z$ starts to dominate and yields a solution like 
\begin{eqnarray}
v_k\sim C_kz.
\end{eqnarray}
Matching both these solutions at the `sound horizon' we get the coefficient $|C_k|^2=(2z^{*2}c_sk)^{-1}$, where asterisk denotes the value at `sound horizon' crossing. The power spectrum of the comoving curvature perturbation at the `sound horizon crossing' can be determined as 
\begin{eqnarray}
{\mathcal P}_{\mathcal R}^*=\frac{k^3}{2\pi^2}|{\mathcal R}_k^*|^2=\frac{k^3}{2\pi^2}|\zeta_k^*|^2=\frac{k^3}{2\pi^2z^{*2}}|v_k^*|^2,
\label{k-inf-power}
\end{eqnarray}
which turns out to be 
\begin{eqnarray}
{\mathcal P}_{\mathcal R}^*=\frac{1}{8\pi^2}\frac{H^2}{\epsilon_0c_sM_{\rm Pl}^2}\left(\frac{k}{k_{\rm P}}\right)^{n_s-1}\equiv A_s\left(\frac{k}{k_{\rm P}}\right)^{n_s-1}.
\end{eqnarray}
Here $k_{\rm P}$ is dubbed as the pivot scale and for PLANCK observation it is chosen to be $k_{\rm P}=0.05$ Mpc$^{-1}$.
The scale dependence of the curvature scalar spectrum is an artefact of the slowly time varying quantities like $H$, $c_s$ and $\epsilon_0$ which take different values when different modes exit the horizon. The scalar spectral index $n_s$ is the measure of the scale dependence of scalar spectrum which depends on the slow-roll parameters as 
\begin{eqnarray}
n_s-1=\frac{d\ln {\mathcal P}_{\mathcal R}}{d\ln k}=-2\epsilon_0-\epsilon_1-s_0.
\end{eqnarray}

The tensor metric perturbations generated during $k-$inflation evolve in the usual way, leaving the horizon when $k=aH$. Hence the tensor power spectrum would have the same form :
\begin{eqnarray}
{\mathcal P}_{T}=\frac{2}{\pi^2}\frac{H^2}{M_{\rm Pl}^2}\left(\frac{k}{k_{\rm P}}\right)^{n_T}\equiv A_T\left(\frac{k}{k_{\rm P}}\right)^{n_T},
\end{eqnarray}
where the tensor spectral index is 
\begin{eqnarray}
n_T=-2\epsilon_0.
\end{eqnarray}
As these primordial tensor perturbations cannot be directly observed at present, due to lack of technological advances, their imprint on CMB polarization as $B$ modes are being measured by observations like PLANCK and BICPE2. The quantity to be measured or constrained in these observations is the tensor-to-scalar ratio $r\equiv A_T/A_s$ which the $k-$inflation scenario predicts as 
\begin{eqnarray}
r=-8c_sn_T.
\end{eqnarray} 
We note that, the canonical slow-roll inflation predicts the tensor-to-scalar ratio as $r=-8n_T$, which is known as the `consistency-relation' of single-field slow-roll inflation. Hence in $k-$inflation this consistency relation of single-field slow roll inflation would be modified rendering these two scenarios observationally distinguishable. 

%=======================================================================================
\section{Classicalization of primordial perturbations during canonical single-field slow-roll inflationary scenario}
\label{csl-inf}In this section we will briefly review how primordial perturbations, both scalar and tensor, become classical on superhorizon scales when their subhorizon quantum dynamics are modified according to a spontaneous collapse mechanism of quantum mechanics, known as Continuous Spontaneous Localization (CSL) model  \cite{Bassi:2012bg}. This mechanism has been developed in previous literature \cite{Martin:2012pea, Das:2013qwa, Das:2014ada} which we will brief here. 

CSL model of quantum mechanics have many attractive features as a collapse model, such as it is applicable to identical particles and its features help collapse of the wavefunction take place in space.  Another attractive feature of CSL model, which makes it more appealing, is its incorporation of amplification property, according to which the collapse parameter $\gamma$ scales as the total mass of the system so that the localization is stronger for larger system. The modified Schr\"odinger equation in CSL model looks like 
\begin{eqnarray}
d\psi_t&=&\left[-\frac{i}{\tilde h}Hdt+\frac{\sqrt{\gamma}}{m_0}\int d\mathbf{x}(M(\mathbf{x})-\langle M(\mathbf{x})\rangle_t)dW_t(\mathbf{x})\right.\nonumber\\
&&\left.-\frac{\gamma}{2m_0^2}\int d\mathbf{x}(M(\mathbf{x})-\langle M(\mathbf{x})\rangle_t)^2dt\right]\psi_t,
\label{csl-eq}
\end{eqnarray}
where the first term on the right hand side is the standard linear term of the Schr\"odinger equation while the next two terms are stochastic and non-linear respectively. Here $W_t(\mathbf x)$ is the Wiener process which encodes the stochastic nature of the evolution, $M(\mathbf{x})$ 
is called the smeared mass density operator and the collapse parameter $\gamma$ is positive and mass-proportional possessing the feature of amplification property. So far the collapse models of quantum mechanics are built up based on phenomenological arguments and an underlying more fundamental theory supporting such collapse dynamics is yet to be developed.\footnote{Adler and collaborators have made attempts  to employ Trace dynamics, which is a classical dynamical theory of noncommuting matrices, as a possible underlying theory of collapse dynamics \cite{Adler:2002fu}.}

As relativistic version of any collapse model of quantum mechanics has not yet been developed, one has to draw analogy from the non-relativistic mechanism of collapse dynamics to implement it in quantum dynamics of the primordial perturbations during inflation. Hence, making analogy with the CSL equation written above,  we add the `CSL-like' non-linear and stochastic terms to the functional Schr\"{o}dinger equation of the gauge-invariant Mukhanov-Sasaki scalar perturbations, $\zeta_{\rm MS}=\delta\varphi+\frac{\dot\varphi}{H}\Phi$, in momentum space as 
\begin{eqnarray}
d\Psi_{\mathbf k}^{\rm R,I}&=&\left[-i\hat{\mathcal{H}}_{\mathbf k}^{\rm R,I}d\tau+\sqrt{\gamma}(\hat{\tilde\zeta}_{\rm{MS}_\mathbf{k}}^{\rm R,I}-\langle\hat{\tilde\zeta}_{\rm{MS}_\mathbf{k}}^{\rm R,I}\rangle)dW_\tau\right.\nonumber\\
&&\left.-\frac\gamma2(\hat{\tilde\zeta}_{\rm{MS}_\mathbf{k}}^{\rm R,I}-\langle\hat{\tilde\zeta}_{\rm{MS}_\mathbf{k}}^{\rm R,I}\rangle)^2d\tau\right]\Psi_{\mathbf k}^{\rm R,I}.
\label{sch-eq-csl}
\end{eqnarray}
Here the Hamiltonian has the form of a simple harmonic oscillator,
\begin{eqnarray}
\hat{\mathcal{H}}_{\mathbf k}^{\rm R,I}=-\frac12\frac{\partial^2}{\partial(\hat{\tilde\zeta}_{\rm{MS}_\mathbf{k}}^{\rm R,I})^2}+\frac12\omega^2(\hat{\tilde\zeta}_{\rm{MS}_\mathbf{k}}^{\rm R,I})^2,
\end{eqnarray}
with a time-dependent frequency $\omega^2(\tau,k)=k^2-a''/a$, due to the expanding background. This functional Schr\"odinger equation has a functional Gaussian as a solution :
\begin{eqnarray}
%\hskip-3cm
&&\Psi_{\mathbf k}^{\rm R,I}(\tau, {\tilde\zeta}_{\rm{MS}_\mathbf{k}}^{\rm R,I})=|\sqrt{N_k}|\exp\left[-\frac{{\rm Re}\,\Omega_k}{2}\left[{\tilde\zeta}_{\rm{MS}_\mathbf{k}}^{\rm R,I}-\bar{{\tilde\zeta}}_{\rm{MS}_\mathbf{k}}^{\rm R,I}\right]^2\right.\nonumber\\
&&\left.+i\sigma_{\mathbf{k}}^{\rm R,I}(\tau)-i\chi_{\mathbf{k}}^{\rm R,I}(\tau){\tilde\zeta}_{\mathbf{k}}^{\rm R,I}-\frac{i{\rm Im}\,\Omega_k}{2}({\tilde\zeta}_{\rm{MS}_\mathbf{k}}^{\rm R,I})^2
\right],
\label{csl-wave}
\end{eqnarray}
where $\bar{{\tilde\zeta}}_{\rm{MS}_\mathbf{k}}^{\rm R,I}$, $\sigma_{\mathbf{k}}^{\rm R,I}(\tau)$ and $\chi_{\mathbf{k}}^{\rm R,I}(\tau)$ are real numbers, $|N_k|=({\rm Re}\,\Omega_k/\pi)^{1/2}$ and $\Omega_k$ satisfies the differential equation as 
\begin{eqnarray}
\Omega_k'=-i\Omega_k^2+i\tilde{\omega}^2(\tau,k),
\end{eqnarray}
with $\tilde\omega^2=\omega^2-2i\gamma$. This parameter $\Omega_k$ plays an important role in determining observables like the power spectrum of the  comoving curvature perturbations ($\mathcal{R}=(H/\dot\varphi)\zeta_{\rm MS}$) :
\begin{eqnarray}
{\mathcal P}_{\mathcal R}=\frac{k^3}{8\pi^2\epsilon_0 M_{\rm Pl}^2}\frac{1}{a^2{\rm Re}\,\Omega_k},
\label{power-spectrum}
\end{eqnarray}
as well as the Wigner function of the mode :
\begin{eqnarray}
&&{\mathcal W}({\tilde\zeta}_{\rm{MS}_{\mathbf k}}^{\rm R}, {\tilde\zeta}_{\rm{MS}_{\mathbf k}}^{\rm I},{\tilde p}_{\rm{MS}_{\mathbf k}}^{\rm R}, {\tilde p}_{\rm{MS}_{\mathbf k}}^{\rm I})=\frac{1}{\pi^2}e^{-{\rm Re}\,\Omega_k({\tilde\zeta}_{\rm{MS}_{\mathbf k}}^{{\rm R}^2}+{\tilde\zeta}_{\rm{MS}_{\mathbf k}}^{{\rm I}^2})}\nonumber\\
&&e^{-\frac{\left({\tilde p}_{\rm{MS}_{\mathbf k}}^{\rm R}+{\rm Im}\,\Omega_k{\tilde\zeta}_{\rm{MS}_{\mathbf k}}^{\rm R}\right)^2}{{\rm Re}\,\Omega_k}}e^{-\frac{\left({\tilde p}_{\rm{MS}_{\mathbf k}}^{\rm I}+{\rm Im}\,\Omega_k{\tilde\zeta}_{\rm{MS}_{\mathbf k}}^{\rm I}\right)^2}{{\rm Re}\,\Omega_k}},
\end{eqnarray}
which can be used as a measure of the quantum or classical nature of the mode. We note that the Wigner function turns out to be a product of four Gaussians, two for field amplitudes and two for its conjugate momentum, where the variance of Gaussians of field amplitude is proportional to the $1/{\rm Re}\,\Omega_k$ and that for momentum is  ${\rm Re}\,\Omega_k$. In a canonical single-field slow-roll inflationary case, without any collapse dynamics, ${\rm Re}\,\Omega_k\rightarrow0$ on superhorizon scales, making the Wigner function highly squeezed in the momentum direction and widely spread in the field direction. Thus, canonical inflationary scenario cannot explain the issue of localization of the initial perturbations in the field variable, as has been observed by several CMBR observations. Hence it is desired to get ${\rm Re}\,\Omega_k\rightarrow\infty$ for superhorizon modes such that the Wigner function squeezes in the field direction in order to explain the localization, hence classicalization, of the primordial perturbations observed so far. Thus we will expect the collapse modified dynamics of inflationary perturbations to squeeze the Wigner function in the field direction. By doing so, the Wigner function would be stretched in the momentum direction before the spontaneous collapse would have taken place. Once the collapse of the wavefunction occurs the mode would be localized in the phase-space, but the value of the momentum of the mode would be random as that is acquired  by a spontaneous collapse. But, that should not raise any concern as there is no observation in the momentum direction of the primordial perturbations. 

The squeezing of the modes, hence classicalization, depends on the collapse parameter $\gamma$ of the model. For a constant collapse parameter $\gamma$, as has been first considered in \cite{Martin:2012pea}, one cannot implement the amplification mechanism of CSL and all the modes, longer or shorter, would become classical with the same rate. Even though, it was observed later in \cite{Das:2013qwa} that a constant $\gamma$ localizes the mode in momentum direction, as the resultant Winger function is squeezed in the momentum direction, and hence fails to explain the classicalization of the primordial scalar modes. It is then proposed in \cite{Das:2013qwa} that the modification of inflationary dynamics with CSL-like modification should also incorporate the essential feature of CSL dynamics, which is the amplification mechanism. It is known from earlier study that the inflationary fluctuations are highly quantum while they are subhorizon and start to become more and more squeezed, and hence classical, when they start to cross the horizon to become superhorizon. Hence it is logical to propose that the collapse of the wavefunctions of the perturbations would take place during or after horizon-crossing and accordingly the collapse parameter $\gamma$, which accounts for the strength of the collapse, should grow with the evolution of the modes. Therefore, a phenomenological form of the collapse parameter is proposed in \cite{Das:2013qwa} as 
\begin{eqnarray}
\gamma=\frac{\gamma_0(k)}{(-k\tau)^\alpha}
\label{gamma}
\end{eqnarray}
where $0<\alpha<2$. 
It was shown in  \cite{Das:2013qwa} that only when $1<\alpha<2$ the Wigner function squeezes in the field direction explaining the localization in the field direction of the primordial perturbations. This shows that incorporating the amplification mechanism does help explain the desired classicalization of the primordial perturbations in certain parameter range. 

The problem of choosing such a collapse parameter is that it would yield a scale-dependent curvature power spectrum which is ruled out by observations. To avoid such discrepancies with observation one can choose 
\begin{eqnarray}
\gamma_0(k)=\tilde\gamma_0\left(\frac{k}{k_0}\right)^{\beta}, 
\label{gamma0}
\end{eqnarray}
where $\tilde\gamma_0$ is a constant and $k_0$ is the comoving wavenumber of the mode which is at the horizon today $k_0=a_0H_0$. This yields the scale dependence of the curvature power as 
\begin{eqnarray}
{\mathcal P}_{\mathcal R}(k)\propto k^{3+\alpha-\beta},
\end{eqnarray}
where $\beta=3+\alpha$ would produce a scale-invariant spectrum satisfying the observations. The form of the curvature power spectrum and scalar spectral index turns out to be \cite{Das:2014ada}
\begin{eqnarray}
{\mathcal P}_{\mathcal R}&=&\frac{1}{8\pi^2 \epsilon_0M_{\rm Pl}}\frac{k_0^2H^2}{\tilde\gamma_0}e^{-(1+\alpha)\Delta N}\left(\frac{k_{\rm P}}{k_0}\right)^{3+\alpha-\beta}\left(\frac{k}{k_{\rm P}}\right)^{n_s-1}\nonumber\\
&\equiv&A_s(k_{\rm P})\left(\frac{k}{k_{\rm P}}\right)^{n_s-1},
\end{eqnarray}
with $\Delta N\sim60$ and where 
\begin{eqnarray}
n_s-1&=&3+\alpha-\beta+2\eta-4\epsilon_0\nonumber\\
&=&\delta-2\epsilon_0-\epsilon_1,
\end{eqnarray}
where we have used $\eta= \epsilon_0-\dot\epsilon_0/(2H\epsilon_0)$ and $\epsilon_1\equiv\dot\epsilon_0/(H\epsilon_0)$, and have defined $\delta\equiv3+\alpha-\beta$. The parameter $\delta$ should be at best of the order of slow-roll parameters, if not zero, to satisfy the observations. 

As the primordial tensor perturbations too are of quantum origin, if they are generated during inflation, one can implement the same collapse dynamics to tensor perturbations and there is no harm in assuming that the same collapse parameter, as in Eq.~(\ref{gamma}), helps make tensor modes classical. As the two polarizations of the tensor modes decouple and each of them acts like a massless scalar, the same analysis one does for scalar perturbations holds good for each polarization mode of the tensor perturbations. In such a case the tensor power spectrum and the tensor spectral index turn out to be  
\begin{eqnarray}
{\mathcal P}_{T}&=&\frac{2}{\pi^2 M_{\rm Pl}}\frac{k_0^2H^2}{\tilde\gamma_0}e^{-(1+\alpha)\Delta N}\left(\frac{k_{\rm P}}{k_0}\right)^{3+\alpha-\beta}\left(\frac{k}{k_{\rm P}}\right)^{n_T}\nonumber\\
&\equiv&A_T(k_{\rm P})\left(\frac{k}{k_{\rm P}}\right)^{n_T},
\label{tensor-power}
\end{eqnarray}
where 
\begin{eqnarray}
n_T=\delta-2\epsilon_0.
\label{tensor-spectral}
\end{eqnarray}
We note that due to collapse dynamics both the scalar and tensor spectral indices are modified by the term $\delta$. If $\beta$ is not exactly equal to $3+\alpha$, which is fair enough to assume, then $\delta$ is non-zero but a constant. Hence, though the spectral indices are modified, the running of the indices would remain unaffected by the collapse dynamics. The consistency relation of canonical single-field slow-roll inflation model does get modified due to presence of collapse dynamics as 
\begin{eqnarray}
r=16\epsilon_0=-8n_T+8\delta.
\end{eqnarray}

%============================================================================================

%========================================================================================
\section{Classicalization of primordial perturbations during $k-$inflationary scenario}
\label{csl-kinf}

In this section we would investigate how CSL-like collapse dynamics makes primordial perturbations classical and modifies the observables in $k-$inflationary scenario. We note that the only difference between the scalar dynamics of slow-roll inflation and $k-$inflation is that the scalar perturbations travel with different speeds in these two scenarios :  while the perturbations in slow-roll inflation travel with the speed of light $(c_s=1)$, the perturbations in $k-$inflation travel with a speed different from speed of light $(c_s\neq1)$. Thus the Schr\"odinger picture analysis of $k-$inflationary scenario would be the same as slow-roll inflation, i.e., the variable $v$ would follow the same functional Schr\"odinger equation with a modification in the frequency as 
\begin{eqnarray}
\omega^2=c_s^2k^2-\frac{a''}{a}.
\label{w-kinf-csl}
\end{eqnarray}

Now, to modify the $k-$inflationary scenario with CSL-like collapse dynamics, one can write the modified Schr\"odinger picture in a similar fashion (as in Eq.~(\ref{sch-eq-csl})) as 
\begin{eqnarray}
d\Psi_{\mathbf k}^{\rm R,I}&=&\left[-i\hat{\mathcal{H}}_{\mathbf k}^{\rm R,I}d\tau+\sqrt{\gamma}(\hat{v}_{\mathbf{k}}^{\rm R,I}-\langle\hat{v}_{\mathbf{k}}^{\rm R,I}\rangle)dW_\tau
\right.\nonumber\\
&&\left.-\frac\gamma2(\hat{v}_{\mathbf{k}}^{\rm R,I}-\langle\hat{v}_{\mathbf{k}}^{\rm R,I}\rangle)^2d\tau\right]\Psi_{\mathbf k}^{\rm R,I}
\end{eqnarray}
where the wavefuctional $\Psi$ is
\begin{eqnarray}
\Psi[v(\tau)]=\sum_{\mathbf k}\Psi_{\mathbf k}^{\rm R}[v_{\mathbf{k}}^{\rm R}(\tau)]\Psi_{\mathbf k}^{\rm I}[v_{\mathbf{k}}^{\rm I}(\tau)],
\end{eqnarray}
and the Hamiltonian $\hat{\mathcal{H}}_{\mathbf k}^{\rm R,I}$ is 
\begin{eqnarray}
\hat{\mathcal{H}}_{\mathbf k}^{\rm R,I}=-\frac12\frac{\partial^2}{\partial(\hat{v}_{\mathbf{k}}^{\rm R,I})^2}+\frac12\omega^2(\hat{v}_{\mathbf{k}}^{\rm R,I})^2,
\end{eqnarray}
with a frequency $\omega$ as given in Eq.~(\ref{w-kinf-csl}). The equation of motion of $v_k$ then would be 
\begin{eqnarray}
v_k''-\tilde\omega^2 v_k=0,
\end{eqnarray}
with $\tilde\omega^2\equiv \omega^2-2i\gamma=c_s^2k^2-a''/a-2i\gamma$. Now, as in the case of slow-roll inflation, we have to phenomenologically choose a form of the collapse parameter $\gamma$ in order to incorporate the amplification mechanism of CSL dynamics. We have earlier noted that the scalar modes in $k-$inflationary model travel with a speed $c_s$ and crosses the sound horizon when $k=aH/c_s$. Thus following similar logic, as given before Eq.~(\ref{gamma}), we propose the form of the collapse parameter for scalar modes generated during $k-$inflation as 
\begin{eqnarray}
\gamma=\frac{\gamma_0(k)}{(-c_sk\tau)^\alpha},
\end{eqnarray}
where $0<\alpha<1$.
Then following the same analysis done in \cite{Das:2013qwa} we found that 
\begin{eqnarray}
{\rm Re}\,\Omega_k=\frac k2(-k\tau)^{1-\alpha}\left(\frac{2\gamma_0(k)}{c_s^\alpha k^2}\right).
\end{eqnarray}
This shows that for the parameter range $1<\alpha<2$ the scalar modes in $k-$inflation becomes classical upon crossing the sound horizon during inflation. 
As ${\rm Re}\,\Omega_k=1/(2|v_k|^2)$, then the comoving curvature power spectrum on superhorizon scales in $k-$inflationary scenario, following  Eq.~(\ref{k-inf-power}), would be 
\begin{eqnarray}
{\mathcal P}_{\mathcal R}=\frac{k^3}{2\pi^2 z^2}|v_k|^2=\frac{k^3}{4\pi^2 z^2{\rm Re}\,\Omega_k},
\end{eqnarray}
where $z$ is as given in Eq.~(\ref{k-inf-z}). This yields, using $a=-1/(H\tau)$, the power spectrum as 
\begin{eqnarray}
{\mathcal P}_{\mathcal R}=\frac{k^2c_s^{\alpha+2} H^2}{8\pi^2\epsilon_0M_{\rm Pl}^2\gamma_0(k)}(-k\tau)^{1+\alpha},
\end{eqnarray}
which is dependent on both time $\tau$ and scale $k$. This unusual feature of dependence on time and scale of the power spectrum, due to introduction of the collapse dynamics, has also been observed previously in \cite{Martin:2012pea, Das:2013qwa}. To get rid of the time dependence of the power spectrum, it was suggested in \cite{Martin:2012pea} to evaluate it at the end of inflation. To do so, we note that 
\begin{eqnarray}
-k\tau=\frac{k}{k_0}e^{-\Delta N},
\end{eqnarray}
where $k_0$ is the comoving wavenumber of the mode which is at the horizon today $(k_0=a_0H_0)$ and for observable modes
$\Delta N\sim50-60$. Along with this we use the form of $\gamma_0(k)$ as given in Eq.~(\ref{gamma0}), in order to obtain a scale-invariant power spectrum. Hence we obtain the form of the comoving curvature power spectrum at the end of inflation as 
\begin{eqnarray}
{\mathcal P}_{\mathcal R}=\frac{k_0^2c_s^{\alpha+2 }H^2}{8\pi^2\epsilon_0M_{\rm Pl}^2\tilde\gamma_0}\left(\frac{k}{k_0}\right)^{3+\alpha-\beta}e^{-(1+\alpha)\Delta N},
\end{eqnarray}
which yields a scale-independent spectrum when $\beta\sim 3+\alpha$. It is then straightforward to determine the scalar spectral index as 
\begin{eqnarray}
n_s-1=\delta-2\epsilon_0-\epsilon_1+(\alpha+2)s_0,
\end{eqnarray}
where $\delta\equiv 3+\alpha-\beta$. Thus the comoving curvature power spectrum can be written in a more formal way as 
\begin{eqnarray}
{\mathcal P}_{\mathcal R}=A_s(k_{\rm P})\left(\frac{k}{k_{\rm P}}\right)^{n_s-1}, 
\end{eqnarray}
where $k_{\rm P}$ is the pivot scale and the amplitude of the scalar spectrum $A_s$ is 
\begin{eqnarray}
A_s(k_{\rm P})=\frac{k_0^2c_s^{\alpha+2 }H^2}{8\pi^2\epsilon_0M_{\rm Pl}^2\tilde\gamma_0}\left(\frac{k_{\rm P}}{k_0}\right)^{3+\alpha-\beta}e^{-(1+\alpha)\Delta N}.
\end{eqnarray}

On the other hand, the tensor perturbation analysis in $k-$inflation would just be the same as in slow-roll inflation as the tensor modes in $k-$inflationary scenario travels with a speed of light as in slow-roll inflation. Hence, while applying the CSL-like collapse dynamics to tensor modes in $k-$inflation we can use the form of $\gamma$ given in Eq.~(\ref{gamma}) and that would yield the same tensor power spectrum and  tensor spectral index as given in Eq.~(\ref{tensor-power}) and Eq.~(\ref{tensor-spectral}) respectively. This would yield the tensor-to-scalar ratio as 
\begin{eqnarray}
r=16\epsilon_0c_s^{-(\alpha+2)}=-8(n_T-\delta)c_s^{-(\alpha+2)}.
\end{eqnarray}
It shows that the collapse dynamics yields a distinct consistency relation for $k-$inflation. 
%==========================================================================================
\section{Discussion}
\label{discussion}

Let us now compare how collapse dynamics change the inflationary observables for both canonical slow-roll and $k-$inflation scenario. First we consider the canonical slow-roll scenario. The observables like scalar spectral index $n_s-1=-2\epsilon_0-\epsilon_1$, the tensor spectral index $n_T=-2\epsilon_0$ and the tensor-to-scalar ratio (along with the consistency relation) $r=16\epsilon_0(=-8n_T)$ in the canonical scenario (without collapse) gets modified as $n_s-1=\delta-2\epsilon_0-\epsilon_1$, $n_T=\delta-2\epsilon_0$ and $r=16\epsilon_0(=-8n_T+8\delta)$ respectively due to incorporating CSL-like collapse dynamics. Here the parameter $\delta\equiv 3+\alpha-\beta$ is a constant, as both the parameters $\alpha$ and $\beta$ which appear in the form of collapse parameter $\gamma$ are constant, and at best can be of the order of slow-roll parameters to be in accordance with observations. It is more reasonable to consider that the parameter $\beta$ does not exactly equal to $3+\alpha$ leading to a non-zero $\delta$ and hence modifying the observables due to collapse dynamics. But, in principle one can have $\beta=3+\alpha$ in which case the collapse dynamics leaves the observables of canonical slow-roll inflation untouched. It is also important to note that as $\delta$ is a constant it leaves the running ($\alpha_s\equiv {\rm d}n_s/{\rm d}\ln k$) and the running of the running ($\beta_s\equiv {\rm d}\alpha_s/{\rm d}\ln k$) of the scalar spectral index unaltered by the collapse dynamics even when $\delta$ is non-zero.

Now let us look at the $k-$inflationary scenario. The observables in generic $k-$inflation case like $n_s-1=-2\epsilon_0-\epsilon_1-s_0$, $n_T=-2\epsilon_0$ and $r=16\epsilon_0c_s(=-8n_Tc_s)$ get modified due to collapse dynamics as $n_s-1=\delta-2\epsilon_0-\epsilon_1+(2+\alpha)s_0$, $n_T=\delta-2\epsilon_0$ and $r=16\epsilon_0c_s^{-(2+\alpha)}\left(=-8(n_T-\delta)c_s^{-(2+\alpha)}\right)$ respectively. We note that the main difference between applying collapse dynamics to canonical slow-roll and $k-$inflation scenario is that even if $\delta=0$ the observables related to scalar perturbations, like $n_s$ and $r$, get modified due to collapse dynamics as non-zero $\alpha$ is required to achieve the classicalization of the perturbations. Even more, unlike the canonical slow-roll inflation scenario, collapse dynamics significantly modify the running and running of the running of scalar spectral indices, as we see now. For $k-$inflation without collapse the running and the running of the running of the spectral indices are
\begin{eqnarray}
\alpha_s&=&-2\epsilon_0\epsilon_1-\epsilon_1\epsilon_2-s_0s_1,\nonumber\\
\beta_s&=&-2\epsilon_0\epsilon_1(\epsilon_1+\epsilon_2)-\epsilon_1\epsilon_2(\epsilon_2+\epsilon_3)-s_0s_1(s_1+s_2),\nonumber\\
\end{eqnarray}
respectively, whereas with the collapse dynamics we get 
\begin{eqnarray}
\alpha_s&=&-2\epsilon_0\epsilon_1-\epsilon_1\epsilon_2+(2+\alpha)s_0s_1,\nonumber\\
\beta_s&=&-2\epsilon_0\epsilon_1(\epsilon_1+\epsilon_2)-\epsilon_1\epsilon_2(\epsilon_2+\epsilon_3)\nonumber\\
&&+(2+\alpha)s_0s_1(s_1+s_2).
\end{eqnarray}
Hence the observables related to the scalar sector of $k-$inflation scenario carry a distinct signature of the collapse dynamics which differ from the generic case. 

The current observations like PLANCK and BICEP2 put constraints on the observables discussed above. 
PLANCK has put tight constraint on $n_s$ and $\alpha_s$ as 
\begin{eqnarray}
n_s&=&0.9645\pm0.0049\nonumber\\
&& (68\%\,{\rm CL}, {\rm PLANCK\,TT,\, TE,\,EE+lowP}),\nonumber\\
\alpha_s&=&-0.0057\pm0.00071\nonumber\\
&&(68\%\,{\rm CL}, {\rm PLANCK\,TT,\, TE,\,EE+lowP}),
\end{eqnarray}
which shows that the data prefers a {\it negative} running of the scalar spectral index. But
when running of the running  of scalar spectral index is allowed to float the recent PLANCK data put constraint on $n_s$, $\alpha_s$ and $\beta_s$ as 
\begin{eqnarray}
n_s&=&0.9569\pm0.0077, \nonumber\\
\alpha_s&=&0.011^{+0.014}_{-0.013},\nonumber\\
\beta_s&=&0.029^{+0.015}_{-0.016},
\end{eqnarray}
at the pivot scale $k_{\rm P}=0.05$ Mpc$^{-1}$ at 68\% CL (PLANCK\,TT,\,TE,\,EE+lowP)\cite{Ade:2015lrj}. Allowing for running of the running of scalar spectral index improves the fit of the temperature spectrum at low multipoles. This indicates that the scalar power spectrum is blue-tilted and favours a {\it positive} running and presence of a running of the running at 2$\sigma$ level along with $\beta_s>\alpha_s$. This observation is also supported by other independent analysis like \cite{Cabass:2016ldu}. Though the generic inflationary models predict {\it negative} $\alpha_s$ and $\beta_s$, it is observed in recent studies that effects of entropy perturbations in multifield scenario \cite{vandeBruck:2016rfv} and some specific model in warm inflationary scenarios \cite{Benetti:2016jhf} can yield {\it positive} $\alpha_s$ and $\beta_s$. On the other hand, the most stringent bound on $r$ comes from combining the data of BICEP2/Keck Array and PLANCK collaboration (a.k.a. BKP) which yields  \cite{Ade:2015lrj}
\begin{eqnarray}
r<0.08,
\end{eqnarray}
at the pivot scale $k_{\rm P}=0.002$ Mpc$^{-1}$ at 95\% CL.
Though PLANCK collaboration do not put any bound on the tensor tilt $n_T$, an independent study \cite{Cabass:2015jwe} on PLANCK+BKP data suggests that the data prefers for a positive tensor tilt 
\begin{eqnarray}
n_T=1.7^{+2.2}_{-2.0}\quad\rm {PLANCK+BKP,}\, \,95\%{\rm CL}.
\end{eqnarray}
Indication of a blue spectrum for primordial tensor modes in the present or future data are also supported by other studies \cite{Kuroyanagi:2014nba,Lasky:2015lej}. But it is difficult to realise a blue-tilted tensor spectrum in the realm of canonical or $k-$inflationary single field models \cite{Cai:2014uka}. Future observations like COrE \cite{Cabass:2015jwe} can yield better resolution in quantifying the tensor-tilt.

%--------------------------------------
\subsection{Increasing sound speed parametrization}
To quantify the observables in CSL-modified $k-$inflationary scenario let us parametrize the slow-roll parameters in terms of e-foldings $N$, as has been suggested by Mukhanov in \cite{Mukhanov:2013tua} :
\begin{eqnarray}
\epsilon_0&=&\frac{\tilde\epsilon_0}{(N+1)^p},
\label{epsilon}\\
c_s&=&\frac{\tilde c_s}{(N+1)^q},
\end{eqnarray}
where $\tilde\epsilon_0$ and $\tilde c_s$ are the values of the respective parameters at the end of inflation which both can be considered as $\mathcal{O}(1)$, $p$ and $q$ are both positive and of order unity. We count the number of e-foldings in a reverse order setting $N=0$ for the end of inflation and thus $dN=\ln a$. Thus in this paramterization the speed of sound increases with time during inflation. This parametrization yields the other required slow-roll parameters as 
\begin{eqnarray}
&&\epsilon_1=\frac{p}{(N+1)},\quad \epsilon_2=\epsilon_3=\frac{1}{(N+1)},\nonumber\\
&&s_0=\frac{q}{(N+1)},\quad s_1=s_2=\frac{1}{(N+1)}.
\end{eqnarray}
One can now write the observables as functions of e-foldings as 
\begin{eqnarray}
n_s-1&=&\delta-\frac{2\tilde\epsilon_0}{(N+1)^p}+\frac{(\alpha+2)q-p}{N+1},\nonumber\\
r&=&\frac{16\tilde\epsilon_0\tilde c_s^{-(\alpha+2)}}{(N+1)^{p-(\alpha+2)q}},\nonumber\\
\alpha_s&=&-\frac{2\tilde\epsilon_0p}{(N+1)^{p+1}}+\frac{(\alpha+2)q-p}{(N+1)^2},\nonumber\\
\beta_s&=&-\frac{2\tilde\epsilon_0p(p+1)}{(N+1)^{p+2}}+\frac{2(\alpha+2)q-2p}{(N+1)^3},\nonumber\\
n_T&=&\delta-\frac{2\tilde\epsilon_0}{(N+1)^p}.
\end{eqnarray}
It is well-known that $n_s-1\sim-2/(N+1)$ matches the recent PLANCK observations well. Hence to quantify the parameters of the theory let us consider the case where $\delta=0$. 
\begin{itemize}
\item First we consider $p=1$. If we consider $\tilde\epsilon_0\sim\tilde c_s\sim 1$ then $(\alpha+2)q\sim1$ yields scalar spectral index in accord with observations.  But this yields $r\sim \mathcal{O}(10)$ for any $N$ which is already ruled out by observations.  
\item If $p>1$ then $p-(\alpha+2)q\sim2$ would yield the correct scalar spectral index. This would again yield $r\sim 16/(N+1)^2$ which for $N\sim 60$ gives $r\sim \mathcal{O}(10^{-3})$ and, both $\alpha_s$ $(\mathcal{O}(10^{-4}))$ and $\beta_s$ $(\mathcal{O}(10^{-5}))$ to be negative.
%\item $p=q=1$ is not a feasible solution as that calls for $\alpha$ to be negative for $n_s\sim0.96$ which does not explain the observed classicality of the modes. 
\begin{itemize}
\item For $q=1$ and $p>1$ we require $\alpha=p-4$ to obtain the observed $n_s$ which gives $5<p<6$ to have $1<\alpha<2$.
\end{itemize}
\end{itemize}
Now let us consider a more general case when $\delta\neq0$.
\begin{itemize}
\item First we consider the case with $p=1$. This yields the observed $n_s$ when $p-(\alpha+2)q\sim \delta(N+1)$. Considering $\tilde\epsilon_0=\tilde c_s=1$ we see that $r\sim 16/N^{\delta(N+1)}$, which means that $\delta(N+1)>1.3$ to obtain $r<0.08$. This would yield negative $\alpha_s$ and $\beta_s$.
\item If $p>1$ then the observed $n_s$ is obtained if $p-(\alpha+2)q=\delta(N+1)+2$ and would satisfy the upper bound on $r$ if $\delta>-0.01$. Hence there is a scope of $\delta$ being negative though very small. In such a case the running at leading order would be $\alpha_s\sim -(\delta(N+1)+2)/(N+1)^2$. If $\delta$ is positive then the running would be negative. But if $\delta$ is negative then also it produces negative $\alpha_s$ for $\delta>-0.01$.
\end{itemize}
%\lipsum[1]
%\begin{widetext}

For the tensor-tilt $n_T$, if $\delta=0$ then we get $n_T\sim-10^{-4} $ for $p=2$. Thus a positive $\delta>10^{-4}$ can yield a blue-tilted tensor spectrum which differs from the generic case.

 Thus we infer from this observation that the collapse dynamics only leaves its imprint in the form of modifying consistency relation and a blue-tilted tensor spectrum when $\delta>0$. Hence unless the consistency relation of slow-roll inflationary era is observationally verified and a tensor-tilt is measured, it would be hard to distinguish the collapse modified inflationary dynamics from its generic scenario. 

%============================
\subsection{Decreasing sound speed parametrization}
Mukhanov's prescription of parametrizing slow-roll parameters \cite{Mukhanov:2013tua} describes a situation where the speed of sound increases during inflation. It is observed in \cite{Khoury:2008wj} that it is also possible to have the varying sound speed decreasing during inflation which will leave distinct features in primordial non-Gaussianity.  Motivated by this idea we propose a parametrization for a decreasing speed of sound during inflation as 
\begin{eqnarray}
c_s=\tilde c_s (N+1)^q,
\label{c-decrease}
\end{eqnarray}
where  $q>0$ and $\tilde c_s$ is the speed of sound at the end of inflation ($N=0$) and is much smaller than unity. We consider the speed of sound at the begining of 60 e-folds of the order of 1 ($c_s(N=60)\lesssim 1$). Thus for $q=1$ we have $\tilde c_s\sim\mathcal{O}(10^{-2})$. This kind of parametrization would yield the slow-roll parameters related to the sound speed as 
\begin{eqnarray}
s_0=-\frac{q}{N+1},\quad s_1=s_2=\frac{1}{N+1}.
\end{eqnarray}
One can then obtain the observables as 
\begin{eqnarray}
n_s-1&=&\delta-\frac{2\tilde\epsilon_0}{(N+1)^p}-\frac{p+q(\alpha+2)}{(N+1)},\nonumber\\
r&=&\frac{16\tilde\epsilon_0c_s^{-(\alpha+2)}}{(N+1)^{p}},\nonumber\\
\alpha_s&=&-\frac{2\tilde\epsilon_0 p}{(N+1)^{p+1}}-\frac{p+q(\alpha+2)}{(N+1)^2},\nonumber\\
\beta_s&=&-\frac{2\tilde\epsilon_0 p(p+1)}{(N+1)^{p+2}}-\frac{2p+2q(\alpha+2)}{(N+1)^3}.
\end{eqnarray}
First we consider the bound on tensor-to-scalar ratio. As per the parametrization $\tilde\epsilon_o\sim c_s(N=60)\sim 1$, we must have $p>1.3$ to satisfy the observational upper bound on $r$. Hence we will consider only the case when $p>1$. 
\begin{itemize}
\item If $\delta=0$, then $p+q(\alpha+2)\sim 2$ would yield the observed $n_s$. But in such a situation one will have $\alpha_s$ and $\beta_s$ to be negative.
\item If $\delta\neq0$, then $p+q(\alpha+2)=\delta(N+1)+2$ would yield the observed $n_s$. In such a case we have $\alpha_s\sim-\delta/(N+1)$ and $\beta_s\sim -2\delta/(N+1)^2$ at leading order. But to have $\alpha_s$ to be positive one requires $\delta(<-0.03)$ and to keep $q>0$ one requires $\delta>-0.01$. Hence it is not possible to get positive $\alpha_s$ and $\beta_s$ even in this paramterization. 
\end{itemize}

It is now trivial to comment that in such paramtrization also we can get a blue-tilted tensor spectrum for positive $\delta$. Hence only blue-tilted tensor-tilt and modified consistency relation for a non-zero positive $\delta$ carry the signature of collapse.

\section{Conclusions}
\label{conclusion}

The issue of quantum-to-classical transition of primordial quantum perturbations generated during inflation still needs attention as, despite having several alternatives to address the issue, like quantum Decoherence, Bohmian mechanics and spontaneous collapse mechanism, it has not yet been possible to converge on the mechanism which might have played role in the early universe. To do so, one must look for concrete observational signatures of the alternatives available in the literature. Among them, the spontaneous collapse mechanism is most viable for leaving distinct observational features as such mechanisms inherently change the standard quantum dynamics, unlike quantum Decoherence and Bohmian mechanics. 

Among the inflationary observables, the scalar spectral index $n_s$ is the one which has been very accurately measured by current PLANCK observations, ruling out the perfect scale-invariant Harrison-Zel'dovich spectrum at $5\sigma$ level. The PLANCK data yields the running of the spectral index $\alpha_s$ to be a {\it negative} quantity, while allowing for the running of the running $\beta_s$ to float the data prefers {\it positive} $\alpha_s$ and $\beta_s$ along with larger $\beta_s$ than $\alpha_s$. Such discrepancy might get resolved by more accurate measurements of these parameters by future observations, like PRISM \cite{Andre:2013afa, Andre:2013nfa}. As the direct detection of CMB  $B-$modes has not yet been made possible, the combined data of PLANCK and BICEP put an upper bound on the tensor-to-scalar ratio $r<0.08$. Also verification of the consistency relation of canonical slow-roll single-field model, which requires independent measurement of $r$ and the tensor tilt $n_T$, awaits for the direct detection of the CMB $B-$modes \cite{Huang:2015gca, Cabass:2015jwe}. A few studies show that the present data prefers a positive $n_T$ i.e. a blue-tilted primordial tensor spectrum \cite{Cabass:2015jwe, Kuroyanagi:2014nba,Lasky:2015lej}.

With this present status of the observational data we look for observational signatures of the collapse modified single-field inflationary model. 
Previously the observational signature of the canonical single-field slow-roll inflationary scenario has been studied in \cite{Das:2014ada} where it was shown that the collapse parameter $\delta$ modifies $n_s$ and $n_T$ along with modifying the consistency relation of the single-field scenario $(r=-8n_T+8\delta)$. It is also noted that $\delta$ being a constant would not affect the running and the running of the running of the scalar and tensor tilts. But as $\delta\equiv 3+\alpha-\beta$ can in principle be zero when $\beta=3+\alpha$, such a case will leave the collapse modified dynamics indistinguishable from the generic case. Hence the collapse modified dynamics is only observationally distinguishable when $\delta$ is non-zero and modifies the consistency relation of the single-field model. Thus verification of the role played by collapse dynamics in the early universe calls for observationally verifying the consistency relation of single-field inflationary models.

In this work we have considered the generalised single-field slow-roll inflation scenario by allowing for non-canonical kinetic term of the inflaton field, known as $k-$inflation. The scenario has been analyzed here by modifying the inflationary mechanism through collapse dynamics. Such an analysis renders more varied observational signatures of collapse dynamics as it modifies $\alpha_s$ and $\beta_s$ along with $n_s$, $n_T$ and the consistency relation of generic $k-$inflation model. As $n_s$, $r$, $\alpha_s$ and $\beta_s$ depend on the collapse parameter $\alpha$ (which is non-zero to explain the observed classicality of the modes), apart from $\delta$, thus in principle this collapse modified dynamics is distinguishable from the generic case even when $\delta$ is identically zero.

To quantify the observational signature of collapse dynamics in primordial observables, we employ two different parametrizations of slow-roll parameters. The one proposed by Mukhanov in \cite{Mukhanov:2013tua} depicts a scenario where the sound speed increases with time during inflation. Along with that we propose another parametrization of the slow-roll parameters in Eq.~(\ref{c-decrease}), motivated by an earlier study \cite{Khoury:2008wj},  where sound speed decreases during the course of inflation. 
Both these parametrizations would render the generic and collapse modified inflationary dynamics (both canonical and $k$-inflation) indistinguishable by predicting the same values of all the observables from the scalar sector for both $\delta=0$ and $\delta\neq0$ cases. 
We note that when $\delta$ is non-zero and positive $\delta>\mathcal{O}(10^{-4})-\mathcal{O}(10^{-2})$, the collapse dynamics yields a blue-tilted tensor spectrum for both canonical and $k-$inflation scenario, where generically one gets a red-tilted tensor spectrum.
Hence one can only distinguish collapse modified dynamics from the generic case if the consistency relation is observationally verified and the tensor tilt is measured iff $\delta$ is non-zero.  

We conclude this discussion by stressing the point that the collapse mechanism in primordial observables is only observationally distinguishable if $\delta\neq0$. If future study succeeds in yielding a relativistic framework for collapse mechanism and proposes $\delta$ to be identically zero, then we infer that it will not leave any observational signature for single-field inflationary models, though it can potentially turn the primordial quantum perturbations classical. A non-zero $\delta$ would leave its imprint in the consistency relation of single-field models and observationally verifying this relation would confirm or rule out spontaneous collapse mechanisms during inflation. A blue-tilted tensor spectrum can potentially hint towards the role of collapse mechanism in the early universe as it is non-trivial to obtain a blue-tilted tensor spectrum in generic canonical or non-canonical scenarios of single-field models \cite{Cai:2014uka}.

%=======================================================================================
%\noindent {\bf Acknowledgments:} 
\begin{acknowledgements}
 Work of S.D. is supported by Department of Science and Technology,  
 Government of India under the Grant Agreement number IFA13-PH-77 (INSPIRE Faculty Award). SK acknowledges the support from the FCT PhD grant SFRH/BD/51980/2012.
This research work is supported by the grant UID/MAT/00212/2013 and COST Action CA15117 (CANTATA).
 \end{acknowledgements}

%\newpage
\label{Bibliography}
\bibliographystyle{h-physrev3}
 \bibliography{k-csl.bib}

\end{document}